\documentclass[twocolumn]{el-author}

\usepackage{cite}

\usepackage[unicode=true,pagebackref=false, bookmarks=false, colorlinks,linkcolor=black,citecolor=red,urlcolor=black,final]{hyperref}
\hypersetup{pdfstartview={XYZ null null null}}

\makeatletter
\def\bstctlcite{\@ifnextchar[{\@bstctlcite}{\@bstctlcite[@auxout]}}
\def\@bstctlcite[#1]#2{\@bsphack
  \@for\@citeb:=#2\do{%
    \edef\@citeb{\expandafter\@firstofone\@citeb}%
    \if@filesw\immediate\write\csname #1\endcsname{\string\citation{\@citeb}}\fi}%
  \@esphack}
\makeatother

\usepackage{caption}
\usepackage{marvosym}

\usepackage{suffix}
\usepackage{mathtools}
\usepackage{amsmath}

\DeclarePairedDelimiterX\MeijerM[3]{\lparen}{\rparen}%
{\,#3\delimsize\vert\begin{smallmatrix}#1 \\ #2\end{smallmatrix}}

\newcommand\MeijerG[8][]{%
  G^{\,#2,#3}_{#4,#5}\MeijerM[#1]{#6}{#7}{#8}}

\WithSuffix\newcommand\MeijerG*[7]{%
  G^{\,#1,#2}_{#3,#4}\MeijerM*{#5}{#6}{#7}}

\newcommand{\ua}{\uparrow}
\newcommand{\nc}{\newcommand}
\nc{\da}{\downarrow} \nc{\hc}{\hat{c}} \nc{\hS}{\hat{S}}
\nc{\bra}{\langle} \nc{\ket}{\rangle} \nc{\eq}{equation (\ref}
\nc{\h}{\hat} \nc{\hT}{\h{T}}\nc{\be}{\begin{eqnarray}}
\nc{\ee}{\end{eqnarray}}\nc{\rd}{\textrm{d}}\nc{\e}{eqnarray}\nc{\hR}{\hat{R}}\nc{\Tr}{\mathrm{Tr}}
\nc{\tS}{\tilde{S}}\nc{\tr}{\mathrm{tr}}\nc{\8}{\infty}\nc{\lgs}{\bra\ua,\phi|}\nc{\rgs}{|\ua,\phi\ket}
\nc{\hU}{\hat{U}}\nc{\lfs}{\bra\phi|}\nc{\rfs}{|\phi\ket}\nc{\hZ}{\hat{Z}}\nc{\hd}{\hat{d}}\nc{\mD}{\mathcal{D}}
\nc{\bd}{\bar{d}}\nc{\bc}{\bar{c}}\nc{\mc}{\mathcal}\nc{\ea}{eqnarray}\nc{\mG}{\mathcal{G}}\nc{\bce}{\begin{center}}
\nc{\ece}{\end{center}}

\begin{document}
\bstctlcite{IEEEexample:BSTcontrol}

\title{On Hybrid-ARQ-Based Intelligent Reflecting Surface-Assisted Communication System}

\author{Y. Ai\textsuperscript{\Letter}, M. Mohamed, L. Kong, A. Al-Samen, M. Cheffena}

\abstract{The intelligent reflecting surface (IRS) is an emerging technique to extend the wireless coverage. In this letter, the performance of hybrid automatic repeat request (hybrid-ARQ) for an IRS-assisted system is analyzed. More specifically, the outage performance of the IRS-aided system using hybrid-ARQ protocol with chase combining, is studied. Asymptotic analysis also show that the outage performance improves better than linearly by increasing number of reflectors of the IRS-aided system. The results also verify the potential of combining the ARQ scheme in the link layer of the IRS-aided system and demonstrate that very small change of path loss condition can impact the performance largely.
 }


\maketitle

\section{Introduction}

Intelligent reflecting surface (IRS) is a revolutionary enabling technique to significantly improve the system performance of wireless system. By functioning as a reconfigurable lens or mirror for the electromagnetic waves and intelligently reconfiguring the radio wave propagation environment, the IRS can improve system coverage especially when line-of-sight (LoS) is not guaranteed \cite{wu2020towards}. To this end, a number of research has been devoted to investigate the performance of IRS under different setups. The ergodic capacity and bit error rate (BER) performance of an IRS-assisted dual-hop UAV communication system are investigated in \cite{yang2020performance}. The analysis on coverage, probability of signal-to-noise ratio (SNR) gain, and delay outage rate of IRS-aided communication system is conducted in \cite{yang2020coverage}. The secrecy performance of IRS-assisted millimeter wave (mmWave) system is studied in \cite{qiao2020secure}.

It is well known that system performance can be further improved with the incorporation of the automatic repeat request (ARQ) scheme in the link layer of the system \cite{ai2016performance}. The \mbox{ARQ} mechanism is a well-established retransmission technique that has been applied in virtually all modern communication systems. The ARQ mechanism can be interpreted as channels with sequential feedback, where the performance can be enhanced by resending data that has been impaired by unfavorable channel conditions with the use of both error correction and error detection codes. Depending on whether the retransmission includes new redundancy bits from the channel encoder or the retransmitted packet is identical to the original transmission, the hybrid ARQ can be further categorized into the following two types: hybrid ARQ with incremental redundancy (IR) and hybrid ARQ with chase combining (CC) \cite{ai2016performance}.

Despite the omnipresence of hybrid ARQ and great potential of IRS, a thorough literature search indicates that the performance of IRS-aided communication with hybrid-ARQ has not been investigated yet to the authors' best knowledge. To fill the gap, we study the outage performance of IRS-aided system with hybrid-ARQ with CC in this letter.

\section{System and channel models}
\begin{figure}[b!]
\centering
  \includegraphics[width=0.62\linewidth,keepaspectratio,angle=0]{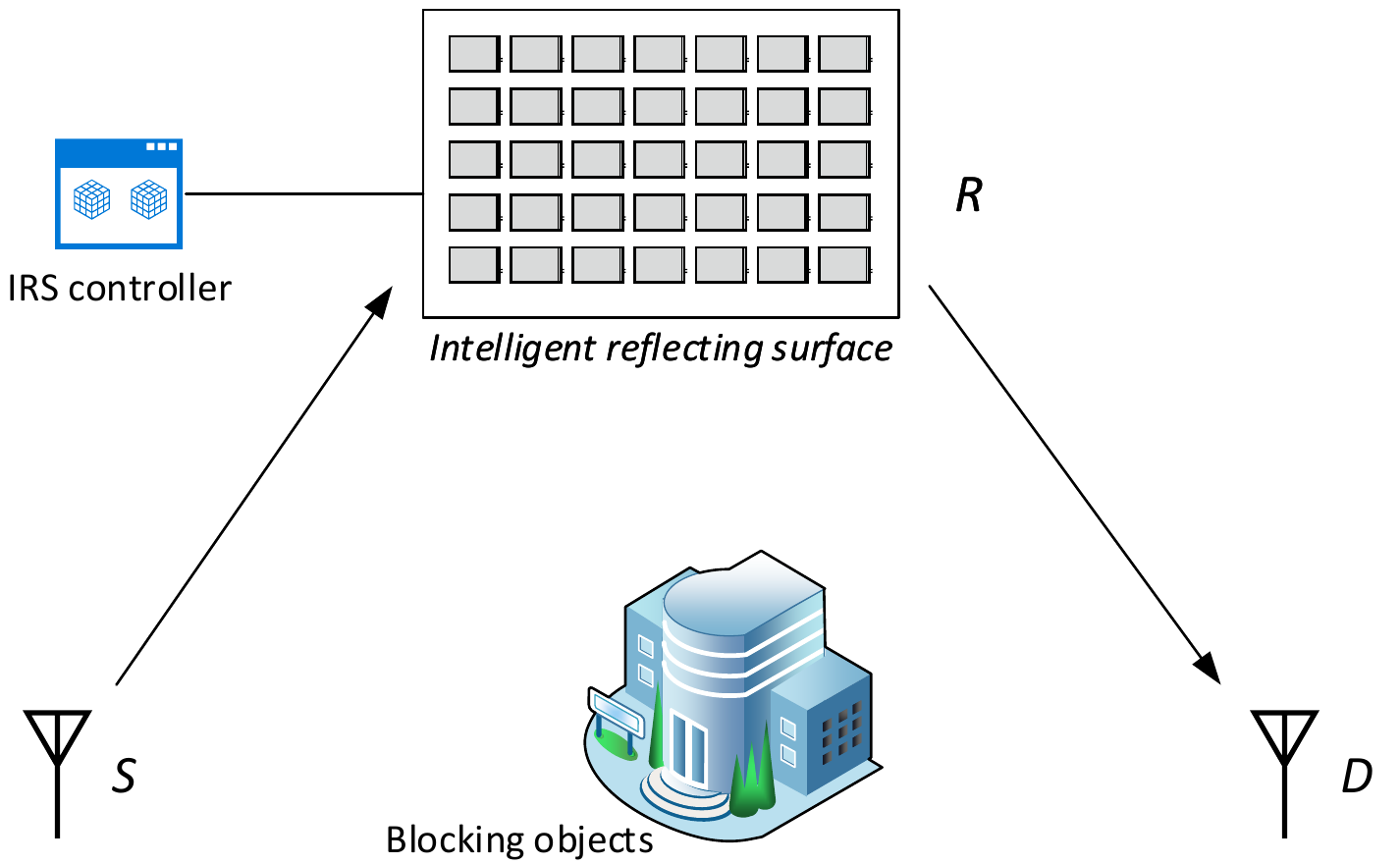}
  \caption{The IRS-aided communication system model.}
  \label{fig:irs_system_model}
\end{figure}

The investigated system model for the IRS-aided communication system is shown in Fig. \ref{fig:irs_system_model}, where a source node \emph{S} tries to communicate with a destination node \emph{D}. There exists no LoS path (i.e., NLoS scenario) between nodes \emph{S} and \emph{D} due to blockage. Instead, an IRS \emph{R} composed of $N$ reflecting elements is placed between \emph{S} and \emph{D} to facilitate the transmission between them. For simplicity, both \emph{S} and \emph{D} are assumed to be equipped with only one antenna. The full channel state information (CSI) of the links \emph{S}-\emph{R} and \emph{R}-\emph{D} are available to the IRS \emph{R} such that maximized SNR at \emph{D} can be achieved. As in \cite{yang2020coverage}, it is also assumed that the signals reflected only once by the IRS are much more significant than the signals reflected multiple times, which are thus ignored. We also consider the case as in \cite{basar2019wireless, atapattu2020reconfigurable, bjornson2019intelligent, yun2020secure} that the reflectors of the IRS are separated large enough such that the elements of the vectors $h_{SR}$ and $h_{RD}$ are independent. Further, Rayleigh fading is considered for the links between the IRS and \emph{D} and between the IRS and \emph{D}. The choice of Rayleigh fading is justified by the fact that in practice LoS between IRS and communicating entities cannot always be guaranteed and scattering from the surrounding environments generally should not be overlooked \cite{basar2020indoor}. Therefore, the RIS-based transmission leads to a double Rayleigh fading distribution from \emph{S} to \emph{D}. It should be noted that when different fading distributions are followed, the central limit theorem (CLT) can still be invoked as in this letter when $N$ is sufficiently large and therefore the main findings obtained in this letter will still be valid.

It should be noted that the IRS can function in the same manner as a relay. However, an IRS and a relay have quite different structures and working principles, which leads to different analysis and performance for them \cite{yang2020performance}. For an IRS, it can be better perceived as part of the so-called smart environment, which simply reflects passively the signals with configuration on the phase to redirect the signals to desired receiver \cite{basar2019wireless}. Instead, a relay need to process the received signals before further forwarding the signals to the desired receiver. A detailed comparison between the IRS-aid and relay-aid communication is done in \cite{di2020reconfigurable}.

For communication with the aid of IRS consisting of $N$ reflecting elements, the received signal at node \emph{D} can be expressed as \cite{yang2020coverage}
\begin{align}
y  =   \sqrt{ P_{s} } \cdot \mathbf{h}_{SR}^{T} \mathbf{\omega}  \mathbf{h}_{RD}    \cdot s + w_{0} ,
 \label{eq:reced_signal}
\end{align}
where $P_{s}$ is the transmit power of $S$, $s$ is the transmitted signal with unit energy, $w_{0}$ is the zero-mean additive white Gaussian noise (AWGN) with variance $N_{0}$, $\mathbf{\omega} = \mathrm{diag}(\varpi_{1}(\phi_{1})e^{j\phi_{1}}, \dots, \varpi_{N}(\phi_{N})e^{j\phi_{N}})$ is the diagonal matrix consisting of the reflection coefficient produced by each reflection element of the IRS, $\mathbf{h}_{SR}$ is the vector with the channel gain from \emph{S} to each element of IRS and the vector $\mathbf{h}_{RD}$ includes the channel gain from each element of IRS to \emph{D}, respectively.

The channel gains for the link between the $l$-th element of IRS and \emph{S} (\emph{D}) can be formulated as \cite{yang2020performance}
\begin{subequations}
\begin{eqnarray}
h_{SR,l} & = \frac{ \alpha_{l} e^{ -j \theta_{l} } }{ d_{1}^{ 0.5n } }, \\
h_{RD,l} & = \frac{ \beta_{l} e^{ -j \varphi_{l} } }{ d_{2}^{ 0.5n } },
 \label{eq:channel_gain}
\end{eqnarray}
\end{subequations}
where $\alpha_{l}$ and $\beta_{l}$ are the amplitudes of the corresponding channel gain and follow independent and identical distributed (i.i.d.) Rayleigh distribution, $\theta_{l}$ and $\varphi_{l}$ are the phases of the corresponding channel gain, $n$ is the path-loss exponent, $d_{1}$ is distances between \emph{S} and \emph{R} and $d_{2}$ is the distance between \emph{R} and \emph{D}. It is assumed that \emph{S} and $D$ are in the far field of the IRS and the IRS is located far enough from \emph{S} and \emph{D} such that the distance between \emph{S} (or \emph{D}) and all elements of IRS can be regarded as the same.

\section{Performance analysis}
Referring to (\ref{eq:reced_signal}), the received instantaneous SNR $\gamma$ at the node \emph{D} can be written as
\begin{align}
\gamma & = \frac{ \left|  \sqrt{ P_{s} } \cdot \mathbf{h}_{SR}^{T} \mathbf{\omega}  \mathbf{h}_{RD}^{T}  \right|^{2} }{ N_{0} }  = \frac{ P_{s} \! \cdot \! \left|   \sum\limits_{l = 1}^{N} \alpha_{l} \beta_{l} \varpi_{l}(\phi_{l})  e^{j( \phi_{l} - \theta_{l} - \varphi_{l} )}  \right|^{2} }{  N_{0} \cdot d_{1}^{n} d_{2}^{n} } .
 \label{eq:reced_snr}
\end{align}

With ideal phase shifting of the IRS (i.e., $\varpi_{i}(\phi_{i}) = 1$ and $\phi_{i} = \theta_{i} + \varphi_{i}$) \footnote{\scriptsize It should be noted that ideal phase shifting require perfect knowledge of channel state information (CSI) at the IRS, which might be challenging. The results obtained under the assumption of ideal phase shifting clearly represent the upper bound of the performance, which provides insights on the potential of IRS-aided performance. An incomplete summary of channel estimation approaches for IRS is given in \cite[Sec.~VI.B]{basar2019wireless}.\\ \phantom{AA} The phase error can potentially come from imperfect phase estimation as well as the use of discrete phase shifters. It is worth mentioning that it is shown in \cite[Fig.~2]{badiu2019communication} that limited quantization bits for the IRS can already provide satisfactory performance improvement. It is also proven in \cite{zhang2020reconfigurable} that the required number of phase shifters to meet some fixed performance target decreases as the number of IRS reflectors grows (e.g., 5 quantization bits already results in very insignificant performance degradation for an IRS of $N = 300$ \cite[Fig.~5]{zhang2020reconfigurable}). This implies that even with imperfect phase shifting, the IRS can potentially provide significant performance improvements. }, the maximum SNR which \emph{D} can be achieved, is given by
\begin{align}
\gamma  = \frac{ P_{s} \cdot \left(   \sum\limits\limits_{l}^{N} \alpha_{l} \beta_{l}  \right)^{2} }{ N_{0} \cdot d_{1}^{n} d_{2}^{n} }  = \overline{\gamma} \cdot  \left(   \sum\limits\limits_{l}^{N} \alpha_{l} \beta_{l}  \right)^{2} ,
 \label{eq:max_reced_snr}
\end{align}
where $ \overline{\gamma} = \frac{ P_{s} }{ N_{0} d_{1}^{n} d_{2}^{n} } $.

With the random variables (RVs) $\alpha_{l}$ and $\beta_{l}$ being i.i.d. Rayleigh distributed with parameter $\frac{1}{\sqrt{2}}$, the CLT can be applied for very large number of reflection elements in the IRS (i.e., $N \gg 1$). In this case, it holds that the RV $(\sum_{l}^{N} \alpha_{l} \beta_{l})$ becomes a Gaussian RV with mean $\frac{N\pi}{4}$ and variance $N \cdot \left( 1 - \frac{\pi^{2}}{ 16 } \right)$. Then, it is straightforward to show that the RV $\gamma$ in (\ref{eq:max_reced_snr}) becomes a noncentral-$\chi^{2}$ distributed RV with the probability density function (PDF) and cumulative distribution function (CDF) given by \cite{yang2020coverage}
\begin{align}
f_{\gamma}(x)  = & \frac{ 1 }{ 2 \sigma^{2} \overline{\gamma} } \cdot \left( \frac{ x }{ \overline{\gamma} \lambda } \right)^{ - \frac{1}{4} } \! \cdot  \exp\!\!\left( - \frac{ x + \lambda \overline{ \gamma }_{1} }{ 2 \overline{\gamma} \sigma^{2} } \right) \cdot I_{ - \frac{1}{2} }\left(  \frac{ \sqrt{ x \lambda } }{ \sqrt{\overline{\gamma} \sigma^{4} } } \right) ,    \label{eq:irs_snr_pdf}  \\
F_{\gamma}(x)  = & 1 - Q_{\frac{1}{2}}\left( \frac{ \sqrt{\lambda} }{ \sigma }, \frac{ \sqrt{ x } }{ \sqrt{ \overline{\gamma} } \sigma } \right)  ,
 \label{eq:irs_snr_cdf}
\end{align}
where $\lambda = \left( \frac{ N \pi }{ 4 } \right)^{2}$, and $ \sigma^{2} = N \cdot \left( 1 - \frac{\pi^{2}}{ 16 }  \right) $, $I_{a}(\cdot)$ is the modified Bessel function of the first class with order $a$, and $Q_{b}(\cdot, \cdot)$ is the generalized Marcum Q-function, which can be efficiently evaluated with Matlab and Mathematica.

\begin{figure}[b]
\centering{\includegraphics[width=0.926\columnwidth]{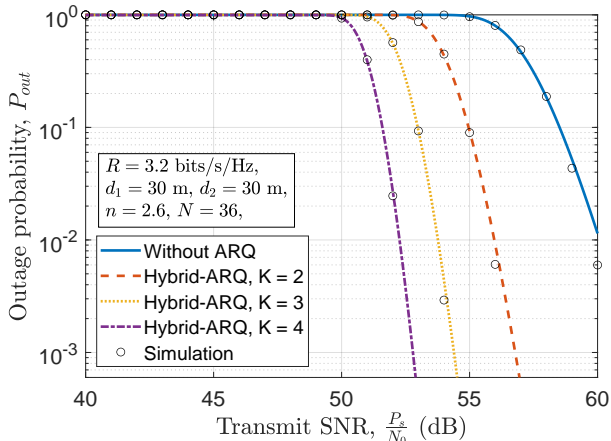}}
\caption{\emph{$P_{out}$ under varying number of ARQ round.}
}\label{pout1}
\end{figure}


For hybrid-ARQ with CC, the packet received in the $k$-th transmission round is combined with the previous received packets and decoding is performed on the combined packet. We assume that information-theoretic capacity achieving channel coding is used. With the optimal combining of the received signals, the mutual information is obtained by combining received SNR over the previous and current transmission rounds \cite{ai2017performance}. The accumulated SNR $\gamma_{K}$ at the node \emph{D} after $K$ ARQ rounds can be written as
\begin{equation}
\gamma_{K} = \sum_{k=1}^{K}  \gamma_{k} =  \sum_{k=1}^{K}  \overline{\gamma}_{k} \cdot  \left(   \sum\limits\limits_{l}^{N} \alpha_{k,l} \beta_{k,l}  \right)^{2},    \label{eq:accumulated_snr}
\end{equation}
where $\gamma_{k}$ is the SNRs for the link from \emph{S} to \emph{D} via IRS at the $k$-th round. The PDF and CDF of the RV $\gamma_{k}$ is given in (\ref{eq:irs_snr_pdf}) and (\ref{eq:irs_snr_cdf}), respectively.

The total accumulated mutual information, $I_{K}$, can be expressed as
\begin{align}
I_{K}  =  \log_{2}( 1 + \gamma_{K})  =  \log_{2}\Bigl( 1 + \sum_{k=1}^{K} \gamma_{k} \Bigr).   \label{eq:accumulated_info}
\end{align}

An outage after $K$ ARQ rounds implies that the accumulated total mutual information $I_{K}$ is still less than the transmission rate $R$. Mathematically, the outage probability $\mathrm{P}_{out}^{(K)}$ after $K$ ARQ rounds can be writen as \cite{mathur2020performance}
\begin{align}
P_{out}^{(K)}  = & \mathrm{Pr}(I_{K} < R) =  \mathrm{Pr}\left( \log_{2}\left( 1 + \sum_{k=1}^{K} \gamma_{k} \right) < R \right)  \nonumber  \\
                     = & \mathrm{Pr}\left( \gamma_{K} < \Theta \right) = F_{ \gamma_{K} }( \Theta ),  \label{eq:outage_prob1}
\end{align}
where $\Theta = 2^{R} - 1$.

Arising from the fact that a noncentral-$\chi^{2}$ RV \mbox{results} from the sum of squares of several i.i.d. Gaussian RVs with non-zero mean; then the sum of several i.i.d. noncentral-$\chi^{2}$ RVs also has a distribution of the same type with the parameters being the sums of the corresponding parameters of the summands. Therefore, the outage probability after $K$ ARQ rounds can be expressed, after some mathematical manipulations, as
\begin{align}
P_{out}^{(K)}  =  1  -  Q_{\frac{K}{2}}\!\!\!\left(    \frac{ \sqrt{NK} \pi }{  \sqrt{ 16 - \pi^{2} }  }  , \left( \frac{ 16 \Theta  }{   N \cdot \left( 16  -  \pi^{2}  \right) \cdot \overline{\gamma}  } \right)^{ \frac{1}{2} } \! \right) .  \label{eq:outage_prob2}
\end{align}

\section{Asymptotic analysis} Next, we consider the system performance when the number of the reflectors $N$ is large.

The following asymptotic expression of generalized Marcum-Q function holds as $q \rightarrow 0$:
\begin{align}
Q_{m}(p, q) \simeq  1 - \frac{ q^{2m} }{ 2^{m} \cdot \Gamma(m) \cdot m } \cdot \exp\Bigl( -\frac{ p^{2} }{ 2 } \Bigr) + o( q^{2m} ).
\label{eq:asy_marcum}
\end{align}

When $N$ is sufficiently large, by utilizing the above asymptotic expression for (\ref{eq:outage_prob2}), we have
\begin{align}
P_{out}^{(K)}  \simeq \frac{ \exp( - \mathcal{C}_{1} N )  }{ N^{ \frac{K}{2} } } \cdot \frac{ C_{2} }{  \overline{\gamma}^{ \frac{K}{2} } }  ,  \label{eq:outage_prob3}
\end{align}
where $\mathcal{C}_{1} = \frac{ K \pi^{2} }{ 2 \cdot ( 16 -  \pi^{2} ) } $ and $\mathcal{C}_{2} = \left( \frac{ 16 \Theta }{ 16 - \pi^{2} } \right)^{ \frac{K}{2} } \cdot \left( 2^{ \frac{K}{2} - 1 } \Gamma( \frac{K}{2} ) \cdot K \right)^{-1} $. The result in (\ref{eq:outage_prob3}) shows that when the reflector number is sufficiently large, the following holds: $P_{out} \propto - a \cdot N - b \cdot \log(N)$, where $P_{out}$ is in $\log$ scale, $a$ and $b$ are some constants related to $K$.

\section{Numerical analysis and discussions}

\begin{figure}[t]
\centering{\includegraphics[width=0.926\columnwidth]{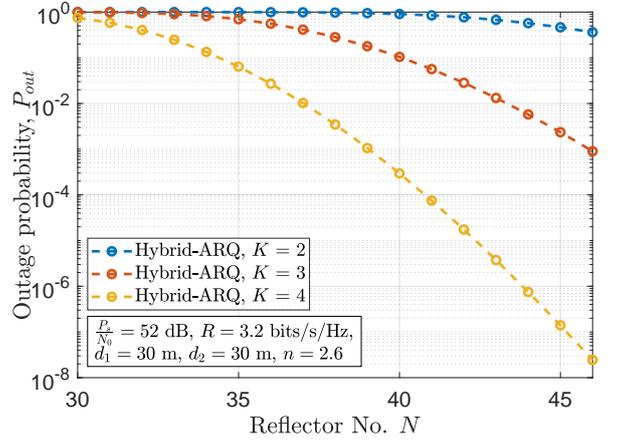}}
\caption{\emph{$P_{out}$ versus varying number of reflecting elements $N$.}
}\label{pout2}
\end{figure}

\begin{figure}[t]
\centering{\includegraphics[width=0.926\columnwidth]{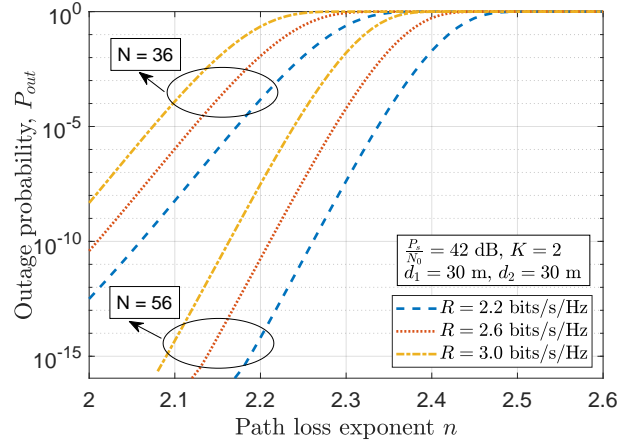}}
\caption{\emph{$P_{out}$ versus varying path loss exponent $n$.}
}\label{pout3}
\end{figure}

Figure \ref{pout1} illustrates the outage performance of the IRS-aided system under different number of ARQ round. It can be seen that compared to the case where ARQ is not utilized, the hybrid ARQ scheme can greatly improve the system performance. However, with the increasing of the ARQ round, the SNR gain becomes smaller. For instance, referring to the outage probability of $10^{-3}$, the transmit SNR is lessened by around 2.4 dB by moving from $K = 2$ to $K = 3$ while the value becomes 1.6 dB when $K$ increases from 3 to 4 rounds. The performance of hybrid ARQ system over IRS-aided system cannot only be enhanced by increasing ARQ rounds, but also by the increase of reflecting elements, as shown in Fig. \ref{pout2}. Fig. \ref{pout3} shows the impact of the path loss exponent on the performance of the ARQ-based IRS-assisted system. It is obvious that the effect of path loss on the performance of IRS-aided system cannot be overlooked. By increasing path loss exponent slightly can lead to relatively significant degradation in performance. This demonstrates that the performance of IRS-aided system is very sensitive to the propagation environment.

\section{Conclusion}
In this paper, we investigate the performance of hybrid ARQ with CC over the IRS-aided communication system. The results represent the best performance that can be achieved with the ARQ-based IRS-aided system under Rayleigh fading. The results clearly verify the potential of combining the ARQ scheme in the link layer of the IRS-aided system as well as demonstrates the sensitivity of propagation environment on the system performance. The potential of the incremental redundancy to support the low-SNR mode is evidenced by the observation that the required transmit SNR to obtain a certain performance target can be lessened significantly as the numbers of reflectors or ARQ rounds increase.

Future work can include the analysis of such system under more general fading channels (e.g., Fisher-Snedecor $\mathcal{F}$ \cite{kong2019intercept}, fluctuating two-ray (FTR) fading \cite{du2020millimeter}, $\alpha$-$\eta$-$\kappa$-$\mu$ fading \cite{ai2020effective}, etc) and under the assumption of imperfect phase shifting \cite{badiu2019communication} at the IRS as well as under the assumption of correlated fading among the reflectors at the IRS.


\vskip5pt

\noindent Y. Ai, M. Mohamed, A. Al-Samen, M. Cheffena (\textit{Faculty of Engineering, Norwegian University of Science and Technology, 2815 Gj{\o}vik, Norway})

\noindent \Letter \, E-mail: yun.ai@ntnu.no

\noindent L. Kong (\textit{Interdisciplinary Centre for Security, Reliability and Trust, The University of Luxembourg, L-1855 Luxembourg})

\vskip3pt

\balance

\bibliographystyle{IEEEtran}

\bibliography{harq_irs_ref1}

\begin{thebibliography}{10}
\providecommand{\url}[1]{#1}
\csname url@samestyle\endcsname
\providecommand{\newblock}{\relax}
\providecommand{\bibinfo}[2]{#2}
\providecommand{\BIBentrySTDinterwordspacing}{\spaceskip=0pt\relax}
\providecommand{\BIBentryALTinterwordstretchfactor}{4}
\providecommand{\BIBentryALTinterwordspacing}{\spaceskip=\fontdimen2\font plus
\BIBentryALTinterwordstretchfactor\fontdimen3\font minus
  \fontdimen4\font\relax}
\providecommand{\BIBforeignlanguage}[2]{{%
\expandafter\ifx\csname l@#1\endcsname\relax
\typeout{** WARNING: IEEEtran.bst: No hyphenation pattern has been}%
\typeout{** loaded for the language `#1'. Using the pattern for}%
\typeout{** the default language instead.}%
\else
\language=\csname l@#1\endcsname
\fi
#2}}
\providecommand{\BIBdecl}{\relax}
\BIBdecl

\bibitem{wu2020towards}
Q.~Wu and R.~Zhang, ``Towards smart and reconfigurable environment: Intelligent
  reflecting surface aided wireless network,'' \emph{IEEE Commun. Mag.},
  vol.~58, no.~1, pp. 106--112, Jan. 2020.

\bibitem{yang2020performance}
L.~Yang, F.~Meng, J.~Zhang, M.~O. Hasna, and M.~Di~Renzo, ``On the performance
  of {RIS}-assisted dual-hop {UAV} communication systems,'' \emph{IEEE Trans.
  Veh. Technol.}, vol.~69, no.~9, pp. 10\,385--10\,390, Sept. 2020.

\bibitem{yang2020coverage}
L.~Yang, Y.~Yang, M.~O. Hasna, and M.-S. Alouini, ``Coverage, probability of
  {SNR} gain, and {DOR} analysis of {RIS}-aided communication systems,''
  \emph{IEEE Wireless Commun. Lett.}, vol.~9, no.~8, Aug. 2020.

\bibitem{qiao2020secure}
J.~Qiao and M.-S. Alouini, ``Secure transmission for intelligent reflecting
  surface-assisted {mmWave} and terahertz systems,'' \emph{arXiv preprint
  arXiv:2005.13451}, 2020.

\bibitem{ai2016performance}
Y.~Ai and M.~Cheffena, ``Performance analysis of {hybrid-ARQ} with chase
  combining over cooperative relay network with asymmetric fading channels,''
  in \emph{Proc. IEEE VTC-Fall}.\hskip 1em plus 0.5em minus 0.4em\relax
  Montreal, Canada: IEEE, Sept. 2016, pp. 1--6.

\bibitem{basar2019wireless}
E.~Basar, M.~Di~Renzo, J.~De~Rosny, M.~Debbah, M.-S. Alouini, and R.~Zhang,
  ``Wireless communications through reconfigurable intelligent surfaces,''
  \emph{IEEE Access}, vol.~7, pp. 116\,753--116\,773, Aug. 2019.

\bibitem{atapattu2020reconfigurable}
S.~Atapattu, R.~Fan, P.~Dharmawansa, G.~Wang, J.~Evans, and T.~A. Tsiftsis,
  ``Reconfigurable intelligent surface assisted two-way communications:
  Performance analysis and optimization,'' \emph{arXiv preprint
  arXiv:2001.07907}, 2020.

\bibitem{bjornson2019intelligent}
E.~Bj{\"o}rnson, {\"O}.~{\"O}zdogan, and E.~G. Larsson, ``Intelligent
  reflecting surface versus decode-and-forward: How large surfaces are needed
  to beat relaying?'' \emph{IEEE Wireless Commun. Lett.}, vol.~9, no.~2, pp.
  244--248, Feb. 2019.

\bibitem{yun2020secure}
Y.~Ai, F.~A.~P. de~Figueiredo, L.~Kong, M.~Cheffena, S.~Chatzinotas, and
  B.~Ottersten, ``Secure vehicular communications through reconfigurable
  intelligent surfaces,'' \emph{arXiv preprint arXiv:2011.14899}, 2020.

\bibitem{basar2020indoor}
E.~Basar, I.~Yildirim, and I.~F. Akyildiz, ``Indoor and outdoor physical
  channel modeling and efficient positioning for reconfigurable intelligent
  surfaces in mmwave bands,'' \emph{arXiv preprint arXiv:2006.02240}, 2020.

\bibitem{di2020reconfigurable}
M.~Di~Renzo, K.~Ntontin, J.~Song, F.~H. Danufane, X.~Qian, F.~Lazarakis,
  J.~De~Rosny, D.-T. Phan-Huy, O.~Simeone, R.~Zhang, M.~Debbah, G.~Lerosey,
  M.~Fink, S.~Tretyakov, and S.~Shamai, ``Reconfigurable intelligent surfaces
  vs. relaying: Differences, similarities, and performance comparison,''
  \emph{IEEE Open J. Commun. Soc.}, vol.~1, pp. 798--807, July 2020.

\bibitem{badiu2019communication}
M.-A. Badiu and J.~P. Coon, ``Communication through a large reflecting surface
  with phase errors,'' \emph{IEEE Wireless Commun. Lett.}, vol.~9, no.~2, pp.
  184--188, Feb. 2019.

\bibitem{zhang2020reconfigurable}
H.~Zhang, B.~Di, L.~Song, and Z.~Han, ``Reconfigurable intelligent surfaces
  assisted communications with limited phase shifts: How many phase shifts are
  enough?'' \emph{IEEE Trans. Veh. Technol.}, vol.~69, no.~4, pp. 4498--4502,
  Apr. 2020.

\bibitem{ai2017performance}
Y.~Ai and M.~Cheffena, ``Performance analysis of {hybrid-ARQ} over full-duplex
  relaying network subject to loop interference under {Nakagami}-$m$ fading
  channels,'' in \emph{Proc. IEEE VTC-Spring}.\hskip 1em plus 0.5em minus
  0.4em\relax Sydney, Australia: IEEE, June 2017, pp. 1--5.

\bibitem{mathur2020performance}
A.~Mathur, Y.~Ai, M.~Cheffena, and M.~R. Bhatnagar, ``Performance of hybrid
  {ARQ} over power line communications channels,'' in \emph{Proc. IEEE
  VTC-Spring}.\hskip 1em plus 0.5em minus 0.4em\relax Antwerp, Belgium: IEEE,
  May 2020, pp. 1--6.

\bibitem{kong2019intercept}
L.~Kong, Y.~Ai, J.~He, N.~Rajatheva, and G.~Kaddoum, ``Intercept probability
  analysis over the cascaded {Fisher-Snedecor} $\mathcal{F}$ fading wiretap
  channels,'' in \emph{Proc. IEEE ISWCS}.\hskip 1em plus 0.5em minus
  0.4em\relax Oulu, Finland: IEEE, Aug. 2019, pp. 672--676.

\bibitem{du2020millimeter}
H.~Du, J.~Zhang, J.~Cheng, Z.~Lu, and B.~Ai, ``Millimeter wave communications
  with reconfigurable intelligent surface: Performance analysis and
  optimization,'' \emph{arXiv preprint arXiv:2003.09090}, 2020.

\bibitem{ai2020effective}
Y.~Ai, A.~Mathur, L.~Kong, and M.~Cheffena, ``Effective throughput analysis of
  $\alpha$-$\eta$-$\kappa$-$\mu$ fading channels,'' \emph{IEEE Access}, vol.~8,
  pp. 57\,363--57\,371, Apr. 2020.

\end{thebibliography}

%
%
%
%
%

\end{document}